\documentclass[conference]{IEEEtran}
\IEEEoverridecommandlockouts
\usepackage{cite}
\usepackage{amsmath,amssymb,amsfonts}
\usepackage{algorithm}
\usepackage{algpseudocode}
\usepackage{amsmath,amssymb}
\DeclareMathOperator*{\argmin}{arg\,min}
\usepackage{graphicx}
\usepackage{textcomp}
\usepackage{xcolor}
\usepackage{algorithm}
\usepackage{graphicx}
\usepackage{graphicx}
\usepackage{wrapfig}
\usepackage{tikz}
\usetikzlibrary{positioning}
\usepackage{pgfplots}
\usepackage{tikz}
\usepackage{algpseudocode}
\usepackage{graphicx} 

\newtheorem{theorem}{Theorem}[section]
\newtheorem{lemma}[theorem]{Lemma}
\def\BibTeX{{\rm B\kern-.05em{\sc i\kern-.025em b}\kern-.08em
		T\kern-.1667em\lower.7ex\hbox{E}\kern-.125emX}}
\begin{document}
	
	\title{TopRank-Based Delivery Rate Optimization for Coded Caching
		under Non-Uniform Demands\\
	}

\author{\IEEEauthorblockN{Mohammadsaber Bahadori}
			\IEEEauthorblockA{\textit{ School of ECE} \\
				\textit{University of Tehran}\\
				Tehran,Iran \\
				m.saberbahadori@ut.ac.ir}
			\and
			\IEEEauthorblockN{Seyed Pooya Shariatpanahi}
				\IEEEauthorblockA{\textit{ School of ECE} \\
					\textit{University of Tehran}\\
					Tehran,Iran \\
					p.shariatpanahi@ut.ac.ir}
				\and
				\IEEEauthorblockN{Behnam Bahrak}
				\IEEEauthorblockA{\textit{TeIAS} \\
					\textit{Khatam University}\\
					Tehran,Iran \\
					b.bahrak@teias.institute}
				
			}
			
			\maketitle
			
			\begin{abstract}
				We study the problem of coded caching with non-uniform file popularity under the setting where the popularity distribution is initially unknown. By reframing the problem, we propose a method inspired by an algorithm from the recommender-systems literature and multi-armed bandits. Unlike prior approaches, which focus on accurately estimating file popularities, our method ranks files relative to one another and partitions them into groups. This perspective is more consistent with the structure of prior approaches as well, since earlier methods also divided files into popular and non-popular groups after estimating their popularities. The proposed approach relies on differences in request counts between files as the basis for ranking, and under many conditions it outperforms the previous algorithm. In particular, we obtain significantly improved performance in scenarios where the number of users in the network is small, the cache storage capacity is limited, or the learning process of the true popularity of files based on observations is contaminated by exploratory or synthetic requests that do not match the true popularity distribution. In these cases, our policy achieves markedly better performance and attains sublinear regret.\end{abstract}
			
			\begin{IEEEkeywords}
				Coded caching, online learning to rank.
			\end{IEEEkeywords}
			
			\section{Introduction}
			Statistics show that currently more than 66\% of the world’s population has access to the Internet and use it \cite{Cisco}. Therefore, it is of great importance to adopt policies and mechanisms for managing Internet networks in order to avoid unnecessary delay and load.
			
			Following the study in \cite{Centralized_Maddahali}, \cite{Decentralized_Maddahali}, which investigated the caching problem from an information-theoretic perspective and presented remarkable results. Various works have focused on adapting this viewpoint to more realistic, practical settings. One important aspect that was not considered in \cite{Centralized_Maddahali} is the non-uniform popularity of files stored at the server. Due to the importance and practical relevance of this issue, extensive research has been conducted in this direction \cite{Non_Maddahali}, \cite{Non_ITA}, \cite{Non-Nikhil}.
			
			In most scenarios, the popularity distribution of files is initially unknown, and it must be learned over time based on observations \cite{Nikhil}. To address this challenge, one can leverage perspectives from bandit algorithms and learning-to-rank techniques \cite{Banditbook}, \cite{RecommSys}. A wide range of problems with different assumptions and characteristics have been studied in these areas, and several creative solutions have been proposed. Some of these problems bear strong similarities to our setting, and with suitable modifications, effective solutions can be derived for our problem.
			
			Paper \cite{Nikhil} addresses the case of non-uniform popularities when no prior information about the popularity distribution is available. In brief, it estimates file popularities based on the history of observations and, using a threshold proportional to the inverse of the product of the number of users and the cache size, partitions files into two groups: popular files (which are cached) and unpopular files. A closer examination reveals several drawbacks of this policy: (a) When the number of requests is small, the estimated popularities are inaccurate. Moreover, if all files are initially assumed to have equal popularity, a long time is required to learn the true popularity of each file. (b) If the local cache size or the number of users is small, or if the number of files is very large and none of the files is significantly more popular than the others, the defined popularity threshold may exceed the popularity of all available files, resulting in no file being cached. (c) If, due to initial searches over all files by users or due to fake and misleading requests for low-popularity files generated by bots or malicious users \cite{fake1}, \cite{fake2}, the observed requests do not follow the true popularity distribution, the algorithm becomes ineffective and may be misled.
			
			Upon deeper inspection, we also realize that precisely estimating the true popularity of files, as attempted by \cite{Nikhil}, is not actually essential. What truly matters is effectively partitioning files into two groups: popular and non-popular.
			In other words, as long as the estimated popular group roughly matches the true popular group, the system functions satisfactorily. For example, if the truly 7th most popular file is estimated as 10th but still included in the popular set, the result remains acceptable. This provides us with a slight degree of flexibility.
			
			Motivated by these conditions and assumptions, we propose an algorithm inspired by \cite{TopRank} that achieves improved performance despite the aforementioned challenges.

			\section{PROBLEM SETTING}
			\subsection{System Model}\label{AA}
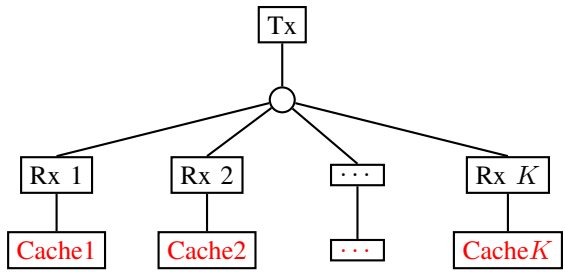
\begin{figure}
				\begin{center}
					\begin{tikzpicture}[node distance=2cm, auto, thick, scale=1, every node/.style={transform shape}]
						
						\node (tx) [draw, rectangle] {Tx};
						\node (shared link) [draw, circle, below of=tx, node distance=1cm]{} ;
						\node (rx1) [draw, rectangle, below of=tx, node distance=2cm, xshift=-3cm] {Rx 1};
						\node (rx2) [draw, rectangle, right of=rx1, node distance=2cm] {Rx 2};
						\node (rx3) [draw, rectangle, right of=rx2, node distance=2cm] {$\dots$};
						\node (rxk) [draw, rectangle, right of=rx3, node distance=2cm] {Rx $K$};
						
						\node (cache1) [draw, rectangle, below of=rx1, node distance=1cm, text=red] {Cache1};
						\node (cache2) [draw, rectangle, below of=rx2, node distance=1cm, text=red] {Cache2};
						\node (cache3) [draw, rectangle, below of=rx3, node distance=1cm, text=red] {$\dots$};
						\node (cachek) [draw, rectangle, below of=rxk, node distance=1cm, text=red] {Cache$K$};
						
						\draw[-] (tx) to node[right] {} (shared link);
						\draw[-] (shared link) -- (rx1.north);
						\draw[-] (shared link) -- (rx2.north);
						\draw[-] (shared link) -- (rx3.north);
						\draw[-] (shared link) -- (rxk.north);
						\draw[-] (rx1) -- (cache1);
						\draw[-] (rx2) -- (cache2);
						\draw[-] (rx3) -- (cache3);
						\draw[-] (rxk) -- (cachek);

					\end{tikzpicture}
				\end{center}
				
				\caption{Network Model}
				\label{Network Model}
			\end{figure}

			We study a system consisting of one server and $K$ users. The server is connected to the users through an error-free link. Inside the server, there are $N$ files, each of size $F$ bits, denoted by $W_1,W_2,\dots,W_N$. Throughout the remainder of this paper, we assume that $[P] := \{1,2, \ldots, P\}$. We denote the popularity of file $i\in[N]$ by $p_i$, and the popularity distribution of all files on the server by $\mathbf{p}=(p_1,p_2,\dots,p_N)$.
			We also assume that the files are sorted according to their popularity, or in other words, $p_1\geq p_2\geq \dots\geq p_N$.
			We assume that in the considered network, each user has a cache of size $MF$ bits, where $M \in[0,N]$. This structure is shown in Fig.~\ref{Network Model}.
			Communication at time $t$ consists of two phases:
			\begin{itemize}
				\item \textbf{Placement Phase}: In this phase, before users request their desired files, the server based on past observations and predictions of future requests, decides how and with what strategy to fill each user’s cache using the files stored on the server. We denote the content placed in the cache of user $k\in [K]$ at time $t$ by $Z^t_k$.
				An important point is that this phase takes place during low-traffic periods of the network. The only limiting factor in this phase is the size of the cache memory.
				Moreover, We also do not consider any rate for transferring data from the server into the cache memories in this phase.
				\item \textbf{Delivery Phase}: In this phase, each user requests one file from the files available on the server. We denote the request of user $k$ at time $t$ by $d_k^t$, and the set of all user requests at this time by $\mathbf{d^t}=\left(d_1^t,d_2^t,\dots,d_K^t\right)$. In this phase, given the content stored in each user’s cache and the broadcast signal $X^t$ transmitted by the server, with total size $R_{total}(t)F$ bits, each user must be able to recover its requested file.
			\end{itemize}
			Our goal in the placement phase is to fill the users’ cache memories in such a way that during peak network traffic, the load on the network is reduced as much as possible, thereby avoiding delays. In other words, we aim to minimize $R_{total}(t)$.
			It is also important to note that these two phases occur continuously over time, and the server always has the opportunity to update or modify the cache contents based on past observations or changes in policy.
			
			Since file popularities differ and storage is limited, files should not be treated uniformly. In \cite{Non_ITA}, a scheme is proposed in which, based on a threshold that depends on the cache size and the number of users ($\frac{1}{KM}$), the files are divided into two groups: popular and non-popular. Files whose popularity exceeds the introduced threshold are considered members of the popular group in the system and are stored using the idea presented in \cite{Decentralized_Maddahali}, known as decentralized coded
			caching. Moreover, in this method, during the placement phase and the caching process, we do not take into account the differences in popularity among the files in the popular group; instead, we assume they have equal popularity and cache them in the same manner. The other files, which fall into the non-popular group, are not stored in any user’s cache and, if requested during the delivery phase, are sent directly from the server.
			
			\subsection{Decentralized Coded Caching}
			
			When the assumption of a centrally coordinated placement phase does not hold, we move toward the decentralized approach \cite{Decentralized_Maddahali}. The main idea presented is that, since users do not have precise information about which other users will be present in the delivery phase, during the placement phase each user, independently of the others, stores a random and uniformly chosen $\frac{M}{N}F$ bit subset of every file in the server in their caches. In the delivery phase, by classifying the requested bits, the server first directly transmits the bits that are not stored in any cache, and for the remaining bits, it exploits coded multicast opportunities by sending coded messages that are simultaneously useful to multiple users.
			
			In \cite{Avestimehr}, the decentralized coded caching problem is also examined for the case in which duplicate requests may occur. Assuming that $N_e(\mathbf{d^t})$ represents the number of distinct requests among $\mathbf{d^t}$, and that we denote by $H_e(\mathbf{d^t})$ the number of distinct requests whose requested file belongs to the popular group and $N^t_2$ also represents the number of popular group files that have been stored in the users’ caches. If we denote by $\mathcal{R}_{D}$ the region in the infinite-dimensional vector space $\left\{R_{D_K}\right\}_{K\in\mathbb{N}}$ (each component of which represents the rate required to satisfy the demands of the $K$ users) obtained through any decentralized prefetching scheme, then, at time $t$, we have:
			\begin{align}
				\label{dfrorm}
				\frac{\mathcal{R}_{D}}{F} =\Bigg\{&\Big\{\frac{R_{D_K}}{F}\Big\}_{K\in\mathbb{N}}\;\Big|\;\frac{R_{D_K}}{F}\geq \nonumber\\ &\quad \mathbb{E}_{\mathbf{d^t}}\left[\frac{N^t_2-M}{M} \left(1-\left(\frac{N^t_2-M}{N^t_2}\right)^{H_{\textup{e}}(\mathbf{d^t})}\right)\right]\Bigg\}
			\end{align}
			
			\subsection{Online Learning}
			
			This learning process is carried out over $n$ rounds. We need a policy for the placement and delivery phases, called $\pi$ that can, over time and based on all observations up to that moment, decide which files are more popular, and which files have a lower probability of being requested in the next round, which means that the contents stored in the caches may change over time and we can update the contents of the caches. 
			In bandit algorithms \cite{Banditbook}, the oracle policy refers to a policy used by an all-knowing agent who is aware of information that is hidden from the learner and always attempts to choose the decision that yields the highest possible reward or the minimum possible loss. In our problem, the oracle seeks to find the minimum rate required to satisfy the users’ demands. To evaluate the performance of a proposed algorithm, we define a quantity called regret, which captures the difference between the incurred cost (or gained reward) and that of the oracle’s policy. The smaller this difference is, the closer the policy 
			$\pi$ is to the oracle. Typically, this regret is defined cumulatively over time. Our overall goal is to achieve sublinear regret, as this implies that, with more observations and experience, the decisions made by the policy $\pi$ get closer to the optimal oracle decisions and the performance improves over time. The mathematical definition of regret over a time horizon $T$ is given as follows:
			\begin{align}
				\mathfrak{R}^{\pi}(T) = \sum_{t=1}^{T}\mathbb{E}\left[R^{\pi}_{total}(t) - R^{O}(t)\right]
			\end{align}
			In the above expression, $\mathfrak{R}^{\pi}(T)$ denotes the regret obtained when using policy $\pi$ over the period $T$, $R^{\pi}_{total}(t)$ denotes the expected rate incurred by the proposed policy $\pi$, and $R^{O}(t)$ denotes the expected rate of the oracle policy at time $t$.
			\section{Proposed Policy}
			
			We propose an algorithm that, in contrast to prior approaches to this problem, makes decisions about the relative popularity ordering of files using concentration inequalities \cite{ConcentrationIneq}, \cite{Freedman}, \cite{self-normalized}. Our proposed algorithm is based on topological sorting: once sufficient evidence about the relative popularity of two files is obtained, this information is stored in a binary relation $G \subseteq [N]^2$, which records all pairwise relationships discovered so far.
			
			More precisely, consider two files $i\in [N]$ and $j\in[N]\backslash \{i\}$. Let the number of requests for file $i$ at time $t$ be denoted by $C_{ti}$, and analogously let $C_{tj}$ denote the number of requests for file $j$. If the cumulative difference of requests over time exceeds a reasonable threshold, we infer with high probability that file $i$ is more popular than file $j$. Equivalently, this implies that $p_i > p_j$. In this case, the pair $(j,i)$ is added to the relation $G$. This comparison and decision process is performed for every pair of files.
			
			After these findings are stored in the relation $G$ at each round, the files are grouped into partitions $\mathcal{P}_{t1}, \dots,\mathcal{P}_{tM_t}$, where $M_t$ denotes the number of nonempty partitions at round $t$. In general, more popular files (or files for which we have not yet obtained sufficient evidence for a precise ranking) are placed in partitions with smaller indices. The partitioning process is performed in a peeling manner; that is, at each stage, all files for which there is no evidence indicating lower popularity compared to the remaining files are grouped into a single partition. This mechanism ensures that within each partition, there are no two files whose relative popularity has been determined. For better intuition, you may refer to lines 17 to 20 of Algorithm~\ref{alg:Algorithm}. Another important point is that, in the proposed algorithm, at time $t$ no two partitions share a common element. Based on all the above explanations, we have $           \sum_{l=1}^{M_t}|\mathcal{P}_{tl}|=N$.
			
			An important point is that if $D_{ti}$ denotes the partition containing file $i$ at time $t$, the algorithm considers the difference in the number of requests between two files only if they belong to the same partition at that time, i.e., $\mathbb{I}\{D_{ti}=D_{tj}\}\left(C_{ti}-C_{tj}\right)$. Taking this difference in the number of requests into account for decision-making, our algorithm is robust and does not get misled when all files are requested simultaneously in an abnormal scenario.
			
			The proposed algorithm for determining the file ordering is based on the algorithm proposed in \cite{TopRank}. In the case where $C_{ti} \in \{0,1\}$, and under the assumptions that $U_{tij} = \mathbb{I}\{D_{ti}=D_{tj}\}\left(C_{ti}-C_{tj}\right)$, $S_{tij} = \sum_{s=1}^t U_{sij}$, and $V_{tij} = \sum_{s=1}^t |U_{sij}|$, with $c \approx 3.43$ and $\delta \in (0,1)$ is a fixed parameter, initially set by us, the following decision threshold is introduced:
			\begin{align}
				S_{tij} \ge \sqrt{2V_{tij}\log\!\big(\tfrac{c}{\delta}\sqrt{V_{tij}}\big)}
			\end{align}
			Although this threshold, with appropriate scaling, remains effective when $C_{ti} \in \{0,K\}$, it is not sufficiently accurate for our setting where $C_{ti} \in [0,K]$, which can be characterized as having a wide range and low variance. To apply the previous threshold under the new conditions, each round must be divided into smaller stages in which the above condition holds. Suppose we define $\theta_t = \max\{C_{t1}, C_{t2}, \dots, C_{tN}\}$. Then, in round $t$, the ranking must be updated $\theta_t$ times while ensuring that condition ($C_{ti} \in \{0,1\}$) holds. In each of these computational steps, each file effectively consumes one of the requests it has received in that round. We repeat this procedure $\theta_t$ times so that all requests are taken into account while the condition remains satisfied. 
			
			However, besides the differences mentioned above between our setting and the problem considered in \cite{TopRank}, in our scenario, all files are displayed to users in every round and all display positions have equal value; therefore, we no longer need to be concerned with managing these aspects.

			As discussed earlier, in each round we must decide, based on the information available up to that point, which files belong to the popular group and should be cached. To this end, we propose two history-based methods.  The main idea of the two proposed methods is to decide how many partitions with the smallest indices (i.e., partitions indexed from 1 up to some index) should be declared as the popular group for the next round. To make this decision, we use the history of requests as a performance metric. Specifically, we assume that the forthcoming requests are highly similar to the last $H$ rounds of requests. In other words, for each possible grouping, we compute the network rate under the assumption that the $\mathcal{D}_H=\left[\mathbf{d^{t-H}}, \dots, \mathbf{d^{t-1}}\right]$ requests correspond to the requests of the next round, and we select the grouping that yields the minimum rate as the popular group. Equivalently, in mathematical terms:  
			\begin{align}
				\label{Fpopg}
				&b^* = \argmin_{b \in \mathbb{N} \cap [1, M_{(t-1)}]} \;\; 
				\bigcup_{i=1}^{b} \mathcal{P}_{(t-1)i}
				\qquad 
				\text{s.t. } \min_{\mathcal{D}_H} R_{total} \nonumber\\
				&N^{t}_2 = \Bigg| \underbrace{\bigcup_{i=1}^{b^*} \mathcal{P}_{(t-1)i}}_{\textbf{The popular group in round $t$}} \Bigg|
			\end{align}

			The two proposed methods are described as follows:
			
			\textbf{A) Method 1:} In this method, we assume that all requests observed over the previous $H$ rounds occur within a single round. Under this assumption, we compute the network rate for different possible groupings, and the grouping that yields the minimum rate is declared as the popular group for the next round. For example, in a two-user network, suppose the requests in the first two rounds were $\left\{1,8\right\}$ and $\left\{3,6\right\}$. To decide for the next round, we assume these two sets occurred together i.e., requests $\left\{1,3,6,8\right\}$ and determine the configuration minimizing the rate.
			An open question is the precise choice of $H$, which can be addressed by empirically evaluating different values and identifying the most suitable one. However, the naive idea of using the entire history can be easily ruled out. The argument is straightforward: under this assumption, after a sufficient number of rounds, it is expected that all files on the server will be requested at least once, which would result in all files being classified as popular which is clearly suboptimal.
			
			\textbf{B) Method 2:} Another approach is again to use the last $H$ rounds of history, but instead of aggregating them into one, we separately compute the rate for each round. We then identify which configuration appears most frequently as optimal across these $H$ rounds, and declare it as the popular set for the next round. The difference between these two methods is that, in the previous method, the computations were performed once for the different cases, whereas in the new method, this process is effectively repeated $H$ times.
			
			All the explanations provided are summarized in Algorithm~\ref{alg:Algorithm}.

			\begin{algorithm}
				\caption{Our Policy (Based on TopRank)}\label{alg:Algorithm}
				\begin{algorithmic}[1] 
					\State $G_0 \gets \emptyset$ \textbf{\&} $ \mathcal{D}_H\gets \emptyset$ \textbf{\&} $c \gets \dfrac{4\sqrt{2/\pi}}{\mathrm{erf}(\sqrt{2})}$ \textbf{\&} $ \gamma\gets -1$ 
					\Statex $M_{0}=1 \Rightarrow \mathcal{P}_{(0)1} \gets [N]$ \textbf{\&} $N_2^1=N$
					\For{$t = 1, \ldots, n$}
					
					\State If $t\neq1$: updating $ \mathcal{D}_H$ and determining the popular  
					\Statex \hspace{\algorithmicindent}group for the current round using (\ref{Fpopg}).
					\State The required information is extracted from $\mathbf{d^{t}}$.
					\State To compute $R_{total}(t)$: the rate associated with requests 
					\Statex \hspace{\algorithmicindent}for the popular group is calculated using (\ref{dfrorm}), while 
					\Statex \hspace{\algorithmicindent}the remaining files are transmitted in full from the\Statex \hspace{\algorithmicindent}server.
					\State $\theta_t=\max\{C_{t1},C_{t2},\dots,C_{tN}\}$ \textbf{\&}  $\forall i\in [N]:\zeta^1_i \gets C_{ti}$
					\For{$ \alpha= 1,2, \ldots, \theta_t$}
					\State $\gamma \gets \gamma+1$
					\State $\forall i\in [N]:\beta_{\gamma i} \gets \mathbb{I}\{\zeta^\alpha_i>0\}$
					\ForAll{$(i,j) \in [N]^2$}
					\State $U_{\gamma ij} \gets \mathbb{I}\{D_{\gamma i}=D_{\gamma j}\}\left(\beta_{\gamma i}-\beta_{\gamma j}\right)$
					\State $S_{\gamma ij} \gets \sum_{s=1}^{\gamma} U_{sij}$ \& $V_{\gamma ij} \gets \sum_{s=1}^{\gamma} |U_{sij}|$
					\vspace{2pt}
					\State $\tau_{\gamma ij}\gets\sqrt{2V_{\gamma ij}\log\!\big(\tfrac{c\sqrt{V_{\gamma ij}}}{\delta}\big)}$
					\EndFor
					\State $G_{\gamma+1} \gets G_\gamma \cup \Big\{ (j,i) :
					S_{\gamma ij} \ge \tau_{\gamma ij}$ \& $V_{\gamma ij}>0 \Big\}$
					\State $m \gets 1$
					\While{$[N] \setminus \bigcup_{c=1}^m \mathcal{P}_{(\gamma +1)c} \neq \emptyset$}
					\State $\mathcal{P}_{(\gamma +1)m} \gets \min_{G_{(\gamma +1)}}\!\left( [N] \setminus \bigcup_{c=1}^{m-1} \mathcal{P}_{(\gamma +1)c} \right)$
					\State $m \gets m + 1$
					\EndWhile
					\State $\forall i\in [N]: \zeta^{\alpha+1}_i\gets \zeta^{\alpha}_i-1$
					\EndFor
					\State The final determined partitions are considered as the 
					\Statex \hspace{\algorithmicindent}partitions at time $t$.
					\EndFor
				\end{algorithmic}
			\end{algorithm}
			
			In the proposed algorithm, the threshold used to determine the file ordering can be made robust against noise-like requests by choosing an appropriate value of $\delta$. Another attractive property of our proposed method is that when multiple files remain in the same partition for a long time, it implies that their popularity levels are nearly identical. This observation is important in practical systems, since, in some cases, even if a specific requested file is unavailable, we can recommend similar files from the same partition as substitutes \cite{Soft}, \cite{similar}.

			\section{Oracle Policy}
			To evaluate the performance of our proposed policy, we need an optimal policy, one that achieves the best possible performance, so that we can compare our policy’s achieved rate with it.
			Regarding this matter, we assume that the oracle possesses two significant capabilities: (a) it knows precisely what user requests are in each round, and (b) it has a list of files sorted by their popularity. Based on the requests in each round, it can only use a marker to indicate, within the list it maintains, from the beginning of the list up to which file constitutes the popular group.

			Considering the two points mentioned above, and also the fact that under the oracle policy each partition has size 1 and there are as many partitions as files, to find the optimal grouping in the oracle policy we proceed similarly to the previous policy, but without the need to evaluate all possible groupings. Since we know exactly which files are requested at each stage, it suffices to compute the rate $N_e(\mathbf{d^t})$ times. Moreover, because we are not dealing with a convex function, the rate must be computed for all distinct requests.

			\begin{lemma}
				\label{increaseN_2}
				\textit{To determine the popular group under the oracle policy, increasing the size of the popular group without adding any new requested files only increases the rate unnecessarily. Intuitively, as the size of the popular group grows, $N^t_2$ increases, while the number of bits stored per file ($\frac{M}{N^t_2}$) decreases. In other words, limited storage capacity is wasted on unneeded files.}
			\end{lemma}
			
			In addition, under a special condition, the following lemma can reduce the number of computations.
			\begin{figure*}[t]
				\centering
				\includegraphics[width=\textwidth]{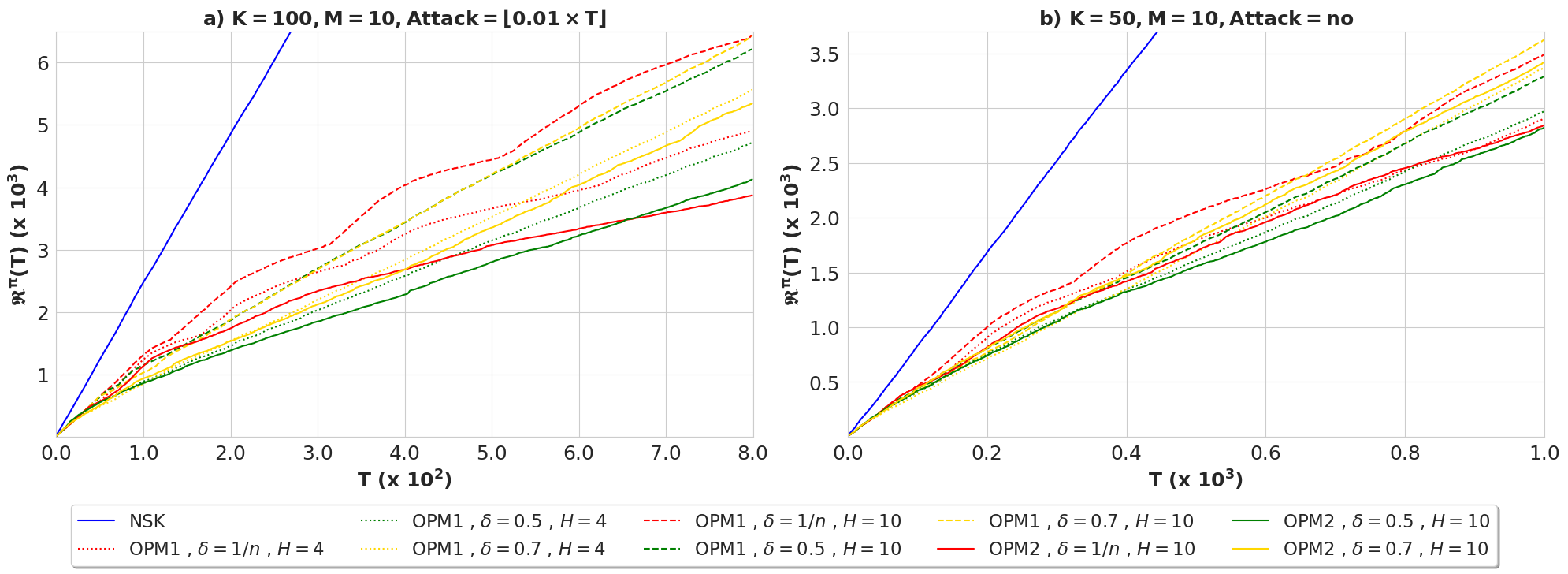}
				\caption{Regret comparison between our proposed policy (OPM1 and OPM2) and the policy in \cite{TopRank} (NSK) under different network settings. The 'NSK' curve continues its approximately linear growth even beyond the domain displayed in the plots.}
				\label{Regret}
			\end{figure*}
			
			\begin{lemma}
				\label{stepsneed}
				\textit{Suppose that at time $t$, during the process of determining the true popular group, it has been established through prior computations that the partition identifying the first $N_2$ files as the popular group, where $h_e(\mathbf{d^t})$ requested files belong to this group, achieves the best performance among the examined candidates. To evaluate the remaining candidates and assess whether increasing the size of the popular group can further improve performance, in the case where $N_2>M$ and $N_2>>Mh_e(\mathbf{d^t})$, the following inequality can be used instead of relying on trial-and-error comparisons.
					\begin{align}
						S_i&<\frac{N_2}{h_e(\mathbf{d^t})}i
					\end{align} 
					In the above expression, $S_i$ denotes the number of files that are added to the current popular group (which currently includes files 1 to file $N_2$). By adding these files, $i$ requested files that were previously members of the non-popular group are now considered part of the popular group. In other words, the popular group becomes $\{1, 2, ..., N_2+S_i\}$. The above inequality indicates that the new grouping is reasonable only if, within fewer than $i \times N_2/h_e(\mathbf{d^t})$ steps, $i$ newly requested files are also included in the popular group.}
			\end{lemma}
			
			An example illustrating how the popular group is computed under the oracle policy is provided in Appendix \ref{Oracle}.

			\section{Numerical Experiment}
			
			In this section, we evaluate the performance of the previously proposed policy in comparison with our own proposed methods. In addition to our proposed policy (with Method 1 and Method 2 reported as 'OPM1' and 'OPM2'), we also implemented the method proposed in \cite{Nikhil}, which is labeled as ‘NSK’ in the results. For the simulations, we use the Movielens 1M dataset\cite{dataset}, which approximately consists of ratings for 4,000 movies, totaling 1 million ratings, collected from 6,000 users. We define the popularity of each file as the total number of ratings it receives divided by the total number of ratings across all files. The requests in each stage are generated via sampling aligned with the empirical popularity distribution of the files.
			
			We implement two different scenarios: one in a network with 100 users experiencing intermittent attacks, and another in a network with 50 users where no unusual requests occur, both under the assumption that $M = 10$. In the 100-user network, we assume that in one round all files are requested from the server, and that this event repeats every 100 rounds. The reason for this atypical request pattern, which we refer to as an “Attack,” may be due to users undertaking an initial exploration to identify all files on the server, or due to the generation of fake requests by malicious users. As shown in the simulations, for fixed $\delta$ and $H$, Method 2 yields lower regret than Method 1, at the cost of increased computations. Also, in Method 1, relying on a long history leads to a noticeable accumulation of estimation error. Furthermore, simulations show that choosing larger values of $\delta$ yields smaller regret in the initial rounds. The reason is that larger $\delta$ reduce sensitivity in ranking, allowing faster differentiation and grouping of files, which results in lower regret compared with the more cautious, accuracy-focused ranking process. However, as time progresses and more accurate rankings become available under smaller $\delta$, the regret gradually decreases. This again demonstrates that in some applications, achieving reasonably fast and moderately accurate grouping may be more important than performing perfectly precise item-level ranking. The figures also shows that selecting an excessively large $\delta$ leads to irrecoverable errors. Hence, the speed of splitting and ranking can be increased only up to a certain point, beyond which errors become unavoidable. 
			
			Other simulation results are provided in Appendix \ref{Simulation}.

			\appendices

			\section{Proof of Lemma ~\ref{increaseN_2}}
			\label{App: increaseN_2}
			We assume that, henceforth, $N^t_2=N_2$, $H_e(\mathbf{d^t})=H_e(d)$, and $N_e(\mathbf{d^t})=N_e(d)$. If $N_2>M$ and $M,H_e(d),N_e(d)$ fixed:
			\begin{align}
				&f =\nonumber\\
				&\frac{N_2-M}{M}\left(1-\left(1-\frac{M}{N_2}\right)^{H_e(d)}\right)+\left[N_e(d)-H_e(d)\right]
			\end{align}
			
			So:
			\begin{align}
				\frac{\partial f}{\partial N_2}  &= \frac{1}{M} - \Bigg[\frac{1}{M}\left(1-\frac{M}{N_2}\right)^{H_e(d)}\nonumber \\ \nonumber &\quad+\frac{N_2-M}{M}\left( H_e(d)\left(1-\frac{M}{N_2}\right)^{H_e(d)-1}\times \frac{M}{N_2^2}\right)\Bigg]\\ 
			\end{align} 
			\begin{align}
				\frac{\partial f}{\partial N_2}  &= \frac{1}{M} - \Bigg[\frac{1}{M}\left(1-\frac{M}{N_2}\right)^{H_e(d)} \nonumber \\ &\quad+\frac{N_2-M}{N_2^2}\times H_e(d)\times \frac{\left(1-M/N_2\right)^{H_e(d)}}{(N_2-M)/N_2}\Bigg]\\ &= \frac{1}{M} - \Bigg[\frac{1}{M}\left(1-\frac{M}{N_2}\right)^{H_e(d)} \nonumber \\  &\quad+\frac{1}{N_2}\left(H_e(d)\left(1-\frac{M}{N_2}\right)^{H_e(d)}\right)\Bigg]\\
				&= \frac{1}{M}(1) - \left[\frac{1}{M}\left(1-\frac{M}{N_2}\right)^{H_e(d)}\left(1+\frac{M H_e(d)}{N_2}\right)\right]\\
				&= \frac{1}{M}\left(1 - \underbrace{\left[\left(1-\frac{M}{N_2}\right)^{H_e(d)}\left(1+\frac{M H_e(d)}{N_2}\right)\right]}_{=g(\frac{M}{N_2})}\right)
			\end{align}
			
			We define $x = \frac{M}{N_2} \in (0,1)$ and $H = H_e(d) > 0$, so that:
			\begin{align}
				g(x)&=(1-x)^H(1+Hx)\\
				g'(x) &= -H(1-x)^{H-1}(1+Hx)+(1-x)^{H}H\\
				&= -H(1-x)^{H-1}(1+Hx-1+x)\\
				&=-H(1-x)^{H-1}(H+1)x < 0 
			\end{align}
			It is straightforward to see that $g'(0)=g'(1)=0$. Combining this observation with $g'(x)<0$, we obtain:
			\begin{align}
				\sup_{x\in(0,1)}g(x) = g(0)=1    
			\end{align}
			By applying this to the main equation, we obtain:
			\begin{align}
				&\frac{\partial f}{\partial N_2} =\nonumber\\
				&\frac{1}{M}\left(1 - \underbrace{\left[\left(1-\frac{M}{N_2}\right)^{H_e(d)}\left(1+\frac{M H_e(d)}{N_2}\right)\right]}_{<1}\right) > 0 
			\end{align}
			We proved that an unjustified increase in the size of the popular group, when it does not cover any additional new requests, leads to an increase in the network rate $\blacksquare$.
			
			\section{Proof of Lemma ~\ref{stepsneed}}
			We assume that, henceforth, $N_2>0$ and $h_e(\mathbf{d^t})>0$. If we denote by $R_{total}^{\{1\}}$ the network rate when considering a popular group of size $N_2$, which under the oracle policy is equivalent to placing files 1 through $N_2$ in the popular group, and by $R_{UP}$ the rate incurred by files that belong to the unpopular group during the delivery phase, we obtain:
			\begin{align}
				&R_{total}^{\{1\}} =\nonumber \\& \frac{N_2-M}{M}\left(1-\left(1-\frac{M}{N_2}\right)^{h_e(\mathbf{d^t})}\right)+R_{UP}
			\end{align}
			Now, suppose that by advancing $S_i$ one step in determining the popular group among the files stored on the server, in addition to increasing the size of the popular group to $N_2+S_i$, we also include $i$ additional files requested by users in the popular group. For this case, we denote the network rate by $R_{total}^{\{2\}}$, and we have:
			\begin{align}
				&R_{total}^{\{2\}} = \nonumber\\&\frac{N_2+S_i-M}{M}\left(1-\left(1-\frac{M}{N_2+S_i}\right)^{h_e(\mathbf{d^t})+i}\right)+R_{UP}-i
			\end{align}
			
			Also, from Appendix \ref{App: increaseN_2}, we know that $S_i$ is the minimum increase in the size of the popular group required to include $i$ new requests among the requests associated with the popular group.
			
			Assuming $N_2>M$ and also $N_2>>Mh_e(\mathbf{d^t})$, we can use the binomial approximation. The basic binomial approximation states that for $|y|<1$ and $|ny|<<1$, $(1+y)^n\approx 1+ny$.
			Applying this approximation to our problem, for the expression appearing in $R_{total}^{\{1\}}$, we obtain:
			\begin{align}
				\left(1-\frac{M}{N_2}\right) ^{h_e(\mathbf{d^t})}\approx 1-h_e(\mathbf{d^t})\frac{M}{N_2}
			\end{align}
			Substituting the derived expression into the initial definition of $R_{total}^{\{1\}}$ yields:
			\begin{align}
				R_{total}^{\{1\}} &\approx \frac{N_2-M}{M}\left(h_e(\mathbf{d^t})\frac{M}{N_2}\right)+R_{UP}\\
				&= h_e(\mathbf{d^t})\left(1-\frac{M}{N_2}\right)+R_{UP}
			\end{align}
			Similarly, for $R_{total}^{\{2\}}$ we can write in the same manner (It is clear that $S_i \geq i$, and we also assume that $N_2+S_i>>M(h_e(\mathbf{d^t})+i)$):
			\begin{align}
				\left(1-\frac{M}{N_2+S_i}\right) ^{h_e(\mathbf{d^t})+i}\approx 1-(h_e(\mathbf{d^t})+i)\frac{M}{N_2+S_i}
			\end{align}
			and by substituting this into the main expression, we obtain:
			\begin{align}
				R_{total}^{\{2\}} &\approx (h_e(\mathbf{d^t})+i)\left(1-\frac{M}{N_2+S_i}\right)+R_{UP}-i
			\end{align}
			Choosing this new group is reasonable only if it leads to a reduction in the network rate. In other words, we must have $R_{total}^{\{2\}} - R_{total}^{\{1\}}<0$. To this end, we first compute the difference between the rates of the two methods:
			\begin{align}
				&R_{total}^{\{2\}} - R_{total}^{\{1\}} \\
				&\approx i-M\left(\frac{h_e(\mathbf{d^t})+i}{N_2+S_i} - \frac{h_e(\mathbf{d^t})}{N_2}\right) -i\\
				& = M\left(-\frac{h_e(\mathbf{d^t})+i}{N_2+S_i} + \frac{h_e(\mathbf{d^t})}{N_2}\right)
			\end{align}
			We now compute the range corresponding to the inequality:
			\begin{align}
				M\left(-\frac{h_e(\mathbf{d^t})+i}{N_2+S_i} + \frac{h_e(\mathbf{d^t})}{N_2}\right)&<0\\
				(N_2+S_i)h_e(\mathbf{d^t})&<N_2(h_e(\mathbf{d^t})+i)\\
				S_i&<\frac{N_2}{h_e(\mathbf{d^t})}i
			\end{align}
			Thus, the claim is proved $\blacksquare$.

			\section{Example for determining the popular group in the oracle policy}
			\label{Oracle}
			For better intuition, suppose the files are indexed according to their popularity. To determine the minimum rate, we sort the requests at each stage in ascending order. Then, for each file in that order, we assume that all files up to that point are stored in the popular set, and we compute the corresponding rate.
			The key point from lemma \ref{increaseN_2} is that we seek the smallest group that contains the newly requested file.
			
			For example, assuming $M=1$, at time $t_0$ consider the request set $\mathbf{d^{t_0}} = \left\{310,434,177,84,165\right\}$. We first sort the requests, i.e., $\{84,165,177,310,434\}$, and define $A_P$ as the subset of requested files included in the popular group, while $A_{UP}$ are those excluded from it. The following example implements the described procedure and computes the rate for all possible configurations:
			\begin{itemize}
				\item $A_p=\{84\}$ \& $A_{up}=\{165,177,310,434\}$ :\\ $R_{total}=\frac{84-1}{1}\left(1-\left(1-\frac{1}{84}\right)^1\right)+\left(5-1\right)=4.9881$
				\item $A_p=\{84,165\}$ \& $A_{up}=\{177,310,434\}$ :\\ $R_{total}=\frac{165-1}{1}\left(1-\left(1-\frac{1}{165}\right)^2\right)+\left(5-2\right)=4.9819$
				\item $A_p=\{84,165,177\}$ \& $A_{up}=\{310,434\}$ :\\ $R_{total}=\frac{177-1}{1}\left(1-\left(1-\frac{1}{177}\right)^3\right)+\left(5-3\right)=4.9662$
				\item $A_p=\{84,165,177,310\}$ \& $A_{up}=\{434\}$ :\\ $R_{total}=\frac{310-1}{1}\left(1-\left(1-\frac{1}{310}\right)^4\right)+\left(5-4\right)=4.9678$
				\item $A_p=\{84,165,177,310,434\}$ \& $A_{up}=\{\}$ :\\ $R_{total}=\frac{434-1}{1}\left(1-\left(1-\frac{1}{434}\right)^5\right)+\left(5-5\right)=\underline{4.9655}$
			\end{itemize}

			In the first case, we assume that only file $\{84\}$ is included in the popular group, meaning the popular group is $\{1,2,\dots ,83,84\}$, and all other files must be fetched directly from the server.
			
			In this example, we also observed that upon encountering a local minimum, one should not stop searching; instead, all possibilities must be examined.
			\begin{figure*}[t]
				\centering
				\includegraphics[width=\textwidth]{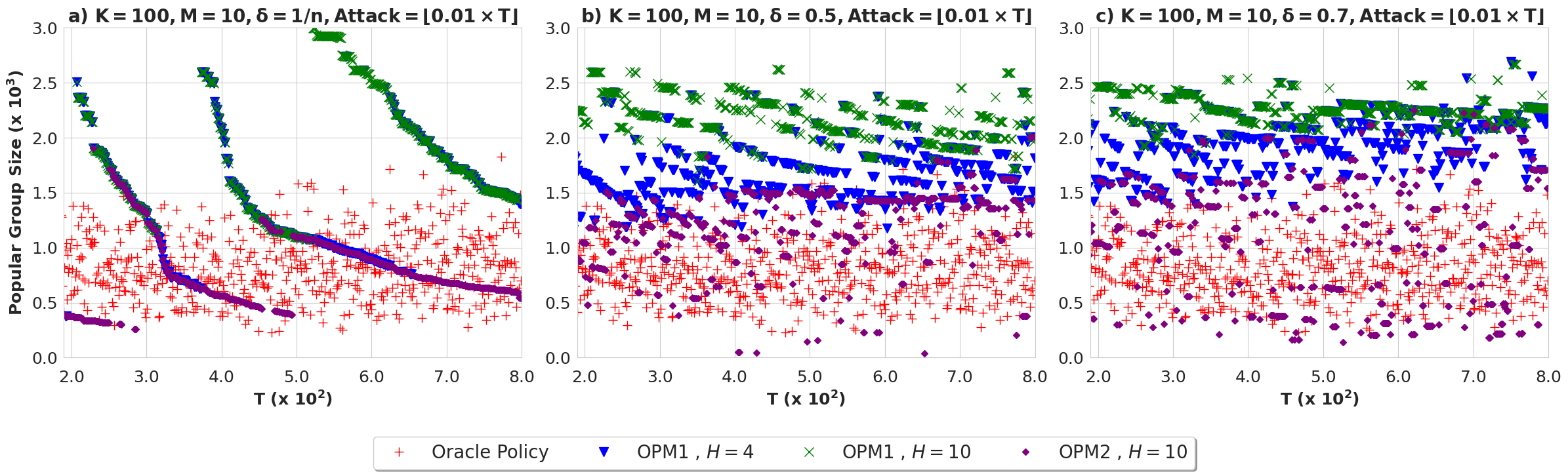}
				\caption{$K=100$ , $M = 10$ , Size of the popular group in different algorithms for a system with parameters $K=100$ and $M = 10$.}
				\label{PopulrGSize}
			\end{figure*}
			\section{Additional Simulation Notes}
			\label{Simulation}
\begin{figure}[htbp]
				
				\centerline{\includegraphics[scale = 0.4]{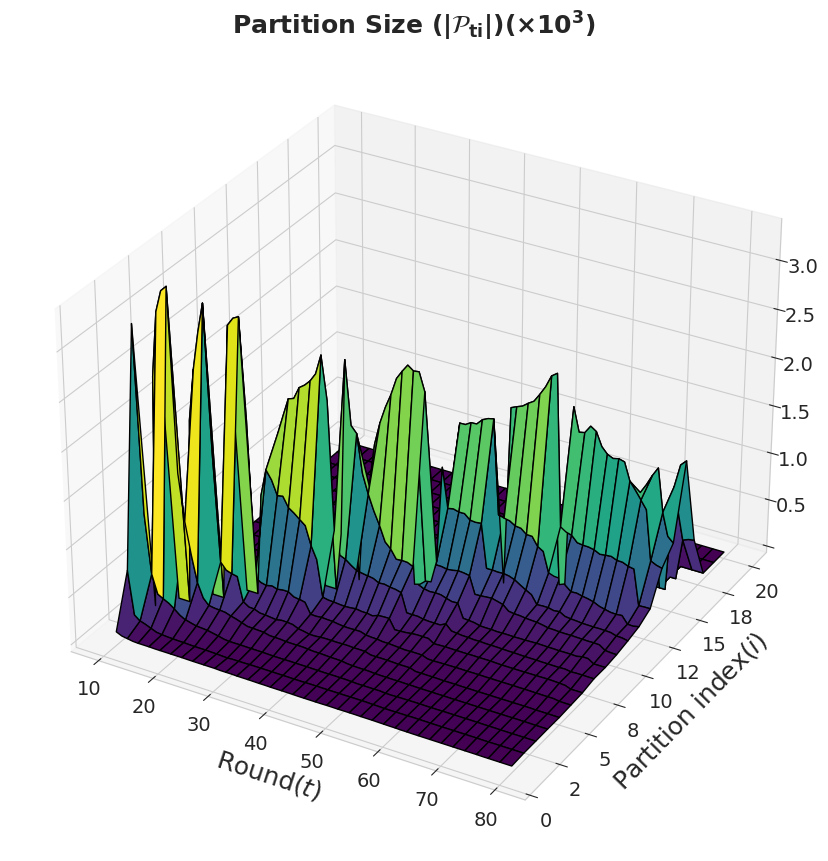}}
				\caption{The sizes of different partitions over time in the 3,000-user system with $\delta=1/(nK)$. This number of users was chosen so that significant changes could be observed within a small time window.}
				\label{Partitionsize}
			\end{figure}

			A very simple and practical observation from implementing the TopRank algorithm \cite{TopRank} is that, as shown in Fig.~\ref{Partitionsize}, more popular files are requested more frequently and separate from other files more quickly. Therefore, one might naively interpret this method as primarily focusing on determining the exact positions of popular files, while less popular files gradually move, like a wave, to partitions with larger indices. This property is important for our problem because, in determining the popular group, we are interested in knowing how many of the top-demand files belong to the popular group. Since we use a trial-and-error approach to answer this question, this behavior allows us not to waste resources on lower-popularity files. Of course, this property is mainly observable in the early and middle rounds; after many rounds, even non-popular files are divided into smaller partitions.
			
			Moreover, as illustrated in Fig.~\ref{PopulrGSize}, the size of the popular group determined by the oracle policy varies from round to round. Without a confidence buffer in identifying the popular group, the system incurs penalties. Such a confidence interval can be generated even using an imprecise ranking. For example, suppose that based on past observations we estimate that the popular group should contain $\{1,2,\dots,7\}$, but our ranking estimates the item popularities as $\{1,2,\dots,9,8,7\}$. If, in the subsequent round, the true popular group becomes larger and our low-precision ranking causes us to cache the wrong files for the upcoming round, we incur less regret. Although this method is not ideal, an interesting point is that our proposed algorithm introduces a reasonable confidence margin. In the proposed algorithm, if two files remain in the same partition for a long time, it indicates that their popularity levels are similar. This feature is particularly useful when determining the popular group because, due to one file, a partition whose members have similar popularity may be included in the popular group, and in the next round, the other members can help reinforce this choice.
			
			In Fig.~\ref{PopulrGSize}, we also observe that if ranking is not done accurately, the members of the popular group remain almost fixed, and the algorithm may mistakenly include many files that were incorrectly placed among the popular files. This again shows that, in Method 1, relying on a long history can lead to deviation from the true request patterns. It is also evident that Method 2 is closer to the actual size of the popular group.
			
		\end{document}